\documentclass[
aps,pra,
twocolumn
]{revtex4}
\usepackage{graphicx}
\usepackage{subfigure}

\usepackage{amsmath}
\usepackage{amssymb}
\usepackage{amsthm}
\usepackage{braket}
\usepackage{color}
\usepackage{microtype}
\usepackage{tikz}
\usepackage{makecell}

\usepackage{hyperref}
\hypersetup{colorlinks,linkcolor={blue},citecolor={red},urlcolor={blue}}

\theoremstyle{plain}

\theoremstyle{plain}

\newenvironment{definition}[1][Definition]{\begin{trivlist}
\item[\hskip \labelsep {\bfseries #1}]}{\end{trivlist}}

\begin{document}

\title{Quantum memory based on concatenating surface codes and quantum Hamming codes}

\author{Menglong Fang}
\author{Daiqin Su}
\email{sudaiqin@hust.edu.cn}

\affiliation{ MOE Key Laboratory of Fundamental Physical Quantities Measurement, Hubei Key Laboratory of Gravitation and Quantum Physics, PGMF, Institute for Quantum Science and Engineering, School of Physics, Huazhong University of Science and Technology, Wuhan 430074, China}


\date{\today}

\begin{abstract}
Designing quantum error correcting codes that promise a high error threshold, low resource overhead and efficient decoding algorithms is crucial to achieve large-scale fault-tolerant quantum computation. The concatenated quantum Hamming code is one of the potential candidates that allows for constant space overhead and efficient decoding. We study the concatenation of surface codes with quantum Hamming codes as a quantum memory, and estimate its error threshold, resource overhead and decoding time. A high error threshold is achieved, which can in principle be pushed up to the threshold of the surface code. 
Furthermore, the concatenated codes can suppress logical errors to a much lower level than the surface codes, under the assumption of comparable amount of resource overhead. The advantage in suppressing errors starts to show for a quantum memory of intermediate scale. Concatenating surface codes with quantum Hamming codes therefore provides a promising avenue to demonstrate small-scale fault-tolerant quantum circuits in the near future, and also paves a way for large-scale fault-tolerant quantum computation.
\end{abstract}

\maketitle

\section{Introduction}

Quantum error correcting codes are introduced to protect fragile quantum information and enable large-scale fault-tolerant quantum computation. The threshold theorem guarantees arbitrarily long computations if the physical error rate remains below a certain threshold~\cite{shor1996fault, aharonov1997fault, knill1998resilient, kitaev2003fault}. A practically favorable quantum error correcting code should have a high error threshold, a fast decoding algorithm, and easy physical implementation. The surface code and its variants can achieve a high error threshold and require only nearest-neighbor gates in experimental implementation, making them the most promising candidates for fault-tolerant quantum computation~\cite{Fowler2012surface}. However, the resource overhead for realizing a practical large-scale fault-tolerant quantum computer based on surface codes is substantial, and fast and accurate decoding algorithms need to be developed. 

Quantum low-density-parity-check (LDPC) codes~\cite{Breuckman2021} are a generalization of the classical LDPC codes and can encode a growing number of logical qubits along with increasing code distance. In particular, the space overhead can be made constant asymptotically, implying a constant code rate. Gottesman's seminal work~\cite{gottesman2014fault} demonstrated that a fault-tolerant error threshold exists for good quantum LDPC codes. Several families of quantum LDPC codes have been developed~\cite{Breukman2016, higgott2023constructions, tillich2013quantum, breuckmann2021balanced, Kovalev2013quantum, panteleev2021degenerate, Lin2024quantum, wang2023abelian, panteleev2022asymptotically, leverrier2022quantum}, and it is believed that quantum LDPC codes with constant overhead and linear code distance will probably be constructed in the near future. The price one needs to pay for using quantum LDPC codes is the requirement of non-local gates, which is challenging for most physical platforms. This challenge could potentially be overcome by designing architectures that accommodate non-local operations. Alternatively, platforms that allow for physical movement of qubits, such as neutral atom and photons, or those with long-range interactions like ion-trapped systems, could also mitigate this issue. Several proposals for implementing quantum LDPC codes in current physical platforms have been studied~\cite{tremblay2022constant, bravyi2024high, xu2024constant, berthusen2024toward}. 

It is recently found that multilevel concatenation of quantum Hamming codes can also achieve constant space overhead asymptotically and allows for efficient decoding~\cite{Yamasaki_2024}. Although concatenated quantum Hamming codes are not quantum LDPC codes, it has been proven that they have an error threshold, and a scheme for fault-tolerant quantum computation has been proposed. 
The circuit-level error threshold and space overhead for fault-tolerant quantum computation based on concatenated quantum Hamming codes and their variants are estimated by decoding level by level and assuming independent error models at each level of concatenation~\cite{yoshida2024concatenate}. However, independent errors in the lowest level of concatenation can generally lead to correlated errors in the higher levels of concatenation. Therefore, a complete simulation of the decoding process is necessary to accurately estimate the error threshold and overhead.   

In this paper, we explore a multi-qubit quantum memory based on the concatenation of surface codes and quantum Hamming codes, which are named surface-Hamming codes. We conduct a comprehensive numerical simulation of the decoding process and estimate the error threshold under the assumption of ideal syndrome extraction and measurement. This approach allows us to account for correlated errors at each level of concatenation. Our numerical results demonstrate the existence of an error threshold for the concatenated quantum Hamming codes. Moreover, by concatenating surface codes with quantum Hamming codes, we observe a significant increase in the error threshold. We further compare the performance of surface-Hamming codes and surface codes, and find that the former is more effective in suppressing logical errors, assuming comparable resource overhead. This highlights the potential advantages of using surface-Hamming codes for fault-tolerant quantum computation. 

The paper is organized as follows. In Sec.~\ref{sec:Hamming} we propose the numerical setup and decoding algorithm, discuss the correlated errors at each level of encoding, and then estimate the error threshold for a quantum memory based on concatenated quantum Hamming codes. In Sec.~\ref{sec:surface-Hamming} we concatenate surface codes with quantum Hamming codes and estimate the corresponding error threshold, and compare the performance of the surface-Hamming codes and surface codes without concatenation. We estimate the decoding time in Sec.~\ref{sec:decoding-time}. Finally, we conclude in Sec.~\ref{sec:conclusion}.

\section{Threshold for Concatenated quantum Hamming codes}\label{sec:Hamming}

It has been recently proven in Ref.~\cite{Yamasaki_2024} that concatenated quantum Hamming codes have an error threshold for fault-tolerant quantum computation. In this section, we consider a quantum memory and use concatenated quantum Hamming codes to protect multiple logical qubits from environmental noise, and estimate the error threshold under the assumption of perfect syndrome extraction and measurement. 

 Quantum Hamming codes~\cite{PhysRevA.54.4741} are Calderbank-Shor-Steane (CSS) codes~\cite{nielsen2010quantum, doi:10.1098/rspa.1996.0136,PhysRevA.54.1098}, which are a family of $[[2^r-1, 2^r-2r-1, 3]]$ codes, where $r$ is a parameter characterizing these codes. 
 The concatenated quantum Hamming codes are constructed through multiple levels of concatenation, with each level comprising several quantum Hamming codes. Every quantum Hamming code in level $l$ is a $[[2^{l+3}-1, 2^{l+3}-2(l+3)-1, 3]]$ code with $l=0, 1, 2\cdots$. We combine all logical qubits encoded in level $l-1$ as a ``register" of level $l$. In particular, the level-$0$ registers are physical qubits. To complete the concatenation in level $l$, we prepare $2^{l+3}-1$ copies of level-$l$ registers to form separate blocks of quantum Hamming code $[[2^{l+3}-1, 2^{l+3}-2(l+3)-1, 3]]$, each of which is constructed by choosing one logical qubit from every level-$l$ register (see Fig.~1 in Ref.~\cite{Yamasaki_2024} for more details).
In the following discussions, we denote the logical qubit at the $i$-th block and the $n$-th level-$l$ register as $\bar{Q}_{n, i}^l$, and the corresponding logical operators are denoted as $\bar{X}_{n, i}^l$ and $\bar{Z}_{n, i}^l$.

\subsection{Error model and syndrome extraction circuit}
\label{sec:circuit}

We consider an error model in which the bit flip and phase flip occur independently, each with the same error probability $p$. It can be described by a quantum channel as
\begin{equation*}
    \mathcal{E} (\rho)= (1-p)^2 \rho +p(1-p)X\rho X+p(1-p)Z\rho Z+p^2 Y\rho Y,
\end{equation*} 
where the Pauli $X$ operator represents a bit flip error, the Pauli $Z$ operator represents a phase flip error, and the Pauli $Y$ operator represents both the bit and phase flips. The error channels are applied only to physical data qubis at the level-0 concatenation.

The quantum Hamming codes are CSS codes; therefore, their stabilizers can be divided into two sets: one containing only tensor products of $X$ operators, and another containing only tensor products of $Z$ operators. Therefore, $X$ errors can be decoded by measuring the stabilizers involving only $Z$ operators, and $Z$ errors can be decoded by measuring the stabilizers involving only $X$ operators. Due to the independence of the decoding process and the equality of the error probability, it is sufficient to study the decoding process of only one type of error. 
The following results are derived solely from the correction of bit flip errors.

\begin{widetext}

    \begin{figure}[htbp]
    \centering
    \includegraphics[width= 0.8\linewidth]{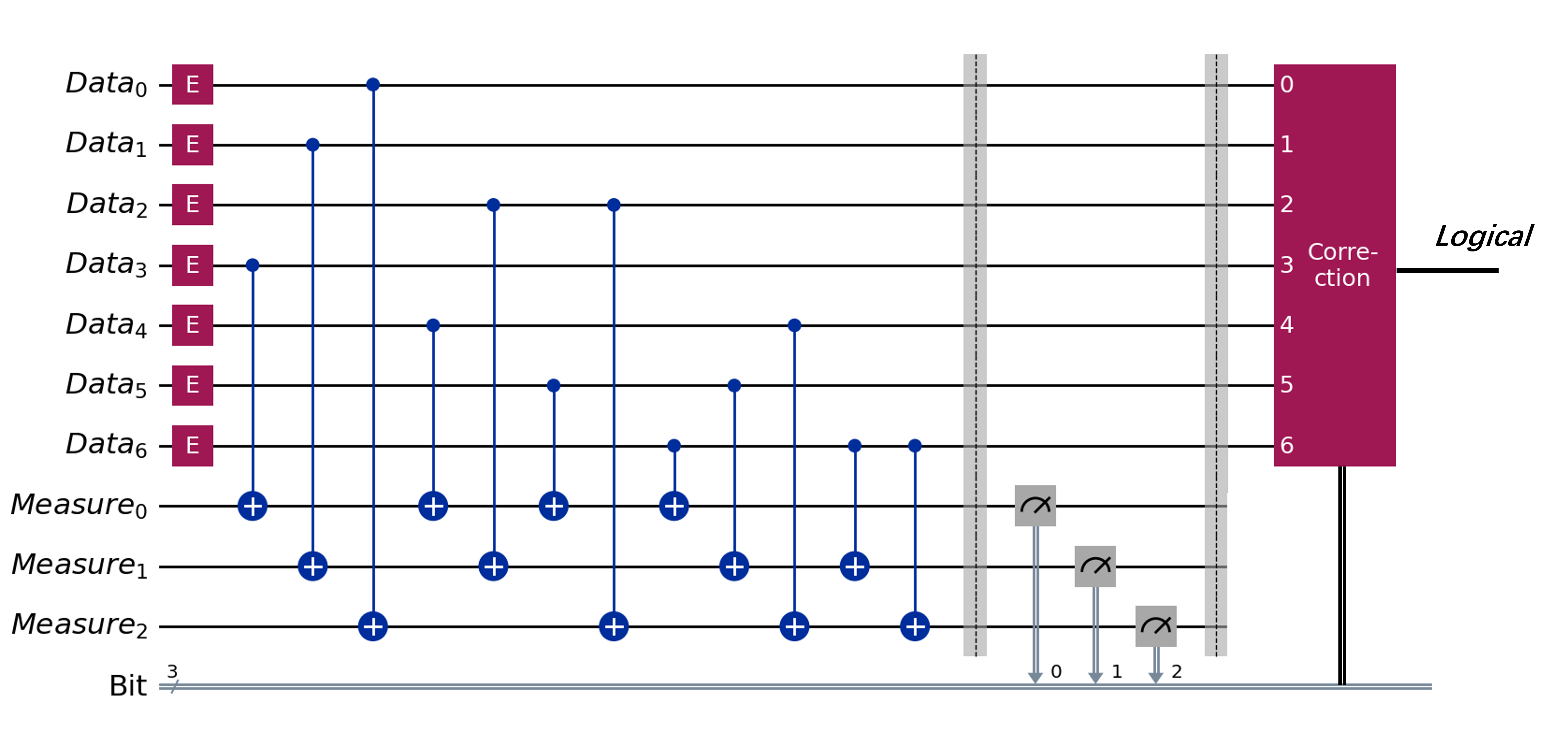}
    \caption{Error correction circuit for a block of the $[[7, 1, 3]]$ quantum Hamming code in our numerical simulation. Errors occur only on seven data qubits. }
    \label{fig:circuit}
\end{figure}
\end{widetext}

In our simulation, we assume that errors occur only on data qubits, while the measurement and syndrome extraction circuits, including the preparation of ancillary (logical) qubits and the application (logical) CNOT gates, are error free. This assumption enables us to estimate the intrinsic error threshold of the concatenated codes. An error correction circuit for a block of the $[[7, 1, 3]]$ quantum Hamming code is shown in Fig.~\ref{fig:circuit}. Three ancillary qubits are initialized in the state $\ket{0}$, interact with seven data qubits through CNOT gates, and are subsequently measured to extract information about errors imposed on the data qubits. Based on results of the syndrome measurement, appropriate recovery operations are applied to the seven data qubits to correct errors and restore the encoded states. The output is a single error-corrected logical qubit of the $[[7, 1, 3]]$ quantum Hamming code, which is then ready to serve as a physical qubit in the next level of concatenation. The error correction circuit for the next level of concatenation can be constructed similarly, and we assume that the ancillary logical qubits are prepared perfectly and the encoded logical CNOT gates are implemented without errors. 

\subsection{Decoder}
\label{sec:decoder}

We use an efficient Hamming-code decoder for the decoding of
concatenated quantum Hamming codes~\cite{nielsen2010quantum}, which executes recursively in a bottom-up manner.
The decoding algorithm begins at level $0$ as follow:
\begin{enumerate}
    \item Measure the syndromes at level $l$. Denote the syndromes of the $n$-th block at level $l$ as $S^{(n)}_l=\{s_i\ |\ i\in[0, k_l-1] \}$, where $k_l=l+3$ is the number of stabilizer generators. Here $s_i \in \{0, 1\}$, and $s_i=1$ when the $i$-th stabilizer generator anticommutes with the error and $s_i=0$ otherwise.
    \item Calculate \begin{equation*}
        N=\sum\limits_{i=0}^{k_l-1} s_i2^i,
    \end{equation*}
    where $N$ is the number that indicates which level-$(l-1)$ logical qubit has suffered an error.
    \item Execute step 2 in parallel on all blocks in level $l$ and correct the error by applying the operator $\bar{X}_{N,n}^{l-1}$ to each block.
    \item If level $l$ is not the final level, repeat the step 1 to 3 at the next level.
\end{enumerate}

The decoding algorithm runs on each block in $O(k_l)$ time, implying that decoding at each level also requires $O(k_l)$ time with parallel execution. When this algorithm operates on a concatenated quantum Hamming code with a total of $\ell$ levels, the overall time complexity is $O(\ell^2)$. 

\subsection{Logical operators} 

In the simulation of the decoding process, the explicit form of each logical operator is needed to identify which logical qubit has suffered an error. These logical operators can be determined by an algorithm introduced in Ref.~\cite{Wilde_2009}. The algorithm 
starts with a set of $m$ Pauli generators $\{g_1, g_2, \cdots, g_m\}$.
\begin{enumerate}
     \item If the generator $g_1$ commutes with all other generators, set it aside in a “set of processed operators”.
    \item If the generator $g_1$ anticommutes with another generator
$g_j$, modify
the remaining generators as follows:
\begin{equation*}
\begin{aligned}
     &  \forall i \in \{2, \cdots, m\}, i\neq j
\\
    g_i\to &\left\{\begin{array}{ccc}
        g_i, &[\;g_i, g_1\,]=0,\ &[\;g_i, g_j\,]=0;   \\
        g_i, &\{g_i, g_1\}=0,\ &\{g_i, g_j\}=0;  \\ 
        g_ig_1,& [\;g_i, g_1\,]=0 ,\ &\{g_i, g_j\}=0;  \\
        g_ig_j,& \{g_i, g_1\}=0,\ &[\;g_i, g_j\,]=0.
    \end{array}\right.
\end{aligned}
\end{equation*}
After the modification, all other generators are commute with both $g_1$ and $g_j$. Then set $g_1$ and $g_j$ aside in the “set of processed operators”.
\item Execute the above procedure recursively to the remaining generators.
\end{enumerate}

By applying this algorithm to  the generator matrix $G$, where each row represents an independent generator of the codewords, we obtain a matrix where each row represents either a logical operator or a stabilizer. After deleting the rows representing the stabilizers, we obtain a matrix denoted as $L$. The generated $i$-th logical operators of the $k$-th block are given by
\begin{equation}
\hat{X}^{l+1}_{i,k}=\bigotimes\limits_j (\bar{X}^{l}_{j,k})^{L_{ij}},\qquad \hat{Z}^{l+1}_{i,k}=\bigotimes\limits_j (\bar{Z}^{l}_{j,k})^{L_{ij}}, 
\end{equation}
where $L_{ij}$ denotes the element of matrix $L$ at row $i$ and column $j$. Here $j\in[1, N_{l}]$, where $N_{l}=2^{l+3}-1$ is the number of level-$l$ registers; $i\in[1, K_{l}]$, where $K_{l}=2^{l+3}-2(l+3)-1$ is the number of logical qubits encoded by a block in level $l$; and $k\in [1, K^{(l-1)}]$ with $K^{(l-1)}=\prod_{i=0}^{l-1} K_i$ the total number of logical qubits in one level-$l$ register. 
Note that $\hat{X}$ and $\hat{Z}$ only label the order of the logical qubits within the same register. At the next level of concatenation,
it needs $(2^{l+4}-1)$ level-$(l+1)$ registers. We should relabel $\hat{X}$ and $\hat{Z}$ to $\bar{X}$ and $\bar{Z}$ at the $r$-th register in the level-$(l+1)$ concatenation, and the result is given by
\begin{equation}
    \bar{X}^{l+1}_{r,(k-1)K_l+i}=\hat{X}^{l+1}_{i,k},\qquad     \bar{Z}^{l+1}_{r,(k-1)K_l+i}=\hat{Z}^{l+1}_{i,k},
\end{equation}

We apply this algorithm to find logical operators of the quantum Hamming code $[[2^n-1, 2^n-2n-1, 3]]$, and obtain the corresponding matrix $L$ for a specific $n$, denoted as $L_n$. For example, the explicit expressions for $n=3$ and $n=4$ are given by
\begin{equation}
L_3=\left(                 
  \begin{array}{ccccccc}   
    0&1&0&1&0&1&0\\  
  \end{array}
\right),               
\end{equation}
\begin{equation}
L_4=\left(                 
  \begin{array}{ccccccccccccccc}   
    1&1&0&1&0&0&0&1&0&0&0&0&0&0&1\\  
    1&0&0&0&0&0&0&0&0&0&0&0&0&1&1\\  
    1&0&1&1&0&0&0&1&0&0&0&0&0&1&0\\  
    0&0&0&1&0&0&0&1&0&0&0&0&1&1&1\\  
    0&0&0&0&1&0&0&1&0&0&0&0&1&0&0\\  
    0&0&1&0&1&1&0&0&0&0&0&0&0&0&0\\  
    1&1&1&0&1&0&1&0&0&0&0&0&1&0&1  
  \end{array}
\right).              
\end{equation}
The expression for $n=5$ is provided in the Appendix~\ref{app:logical-operator}. 

During the process of error correction, we first measure the stabilizers to obtain the syndrome $S$, then apply a decoding algorithm to identify the error $E$, and finally apply a correction operator $Q$ to recover the code state. 
In a real experiment, one cannot detect whether a logical error has occurred or not because the correct code state and erroneous code state share the same syndrome.
However, to estimate the logical error rate using the Monte Carlo simulation, it is necessary to identify which logical qubit has suffered a logical bit-flip error. 
Though the error $E$ is randomly generated in each round of simulation, we know it exactly.
We can check whether the product of the deduced correction operator $Q$ and the error $E$ anticommutes with the logical operator $\bar{Z_i}$. If they anti-commute, then the $i$-th logical qubit has suffered a bit-flip logical error; otherwise, no error has occurred.

\subsection{Correlated errors}
\label{sec:corrlated error}

When a quantum code encodes multiple logical qubits, it is expected that the logical errors for different logical qubits are generally correlated, even though the errors for the physical qubits are independent.
In particular, when the correction operator $Q$ combined with the error $E$ anticommutes with more than one logical operator, the corresponding logical qubits suffer logical errors in a correlated manner. Since the decoding algorithm is performed recursively level by level, those correlated errors can impact the results of subsequent levels. 

To evaluate the correlations between different logical qubits, we compute the Pearson correlation coefficients for all pairs of qubits for the [[15, 7, 3]] quantum Hamming code.
The Pearson correlation coefficient between two random variables $X$ and $Y$ is defined as 
\begin{eqnarray}
    \rho_{X, Y} = \frac{E[(X - \mu_X)(Y - \mu_Y)]}{\sigma_X  \sigma_Y},
\end{eqnarray}
where $\mu_X = E(X)$ and $\mu_Y = E(Y)$ are expectation values, and $\sigma_X^2 = E[(X - \mu_X)^2]$ and $\sigma_Y^2 = E[(Y - \mu_Y)^2]$ are variances. Define a random variable $n_i$, and $n_i=1$ when the $i$-th logical qubit has suffered a logical error and $n_i=0$ otherwise. Consider the correlation between the $i$-th and $j$-th logical qubits, namely, $X = n_i$, $Y = n_j$; and assume that $L_i$ represents the event $n_i=1$. It is straightforward to show that 
\begin{eqnarray}
    \mu_{n_i} &=& P(L_i), ~~~ 
    \mu_{n_j} = P(L_j), \nonumber\\
    \sigma^2_{n_i} &=& 
    P(L_i) - P^2(L_i), ~~~ 
    \sigma^2_{n_j} = 
    P(L_j) - P^2(L_j), 
\end{eqnarray}
therefore we have 
\begin{eqnarray}
    \rho_{n_i, n_j} &=& \frac{P(L_i, L_j) - P(L_i)P(L_j)}{\sqrt{P(L_i)P(L_j)[1 - P(L_i)][1 - P(L_j)]}}. 
\end{eqnarray}
If the logical errors are uncorrelated, then $P(L_i, L_j) = P(L_i) P(L_j)$, so that $\rho_{n_i, n_j} = 0$. A nonzero value of $\rho_{n_i, n_j}$ implies the existence of some sort of correlation.

\begin{figure}[htbp]
    \centering
    \includegraphics[width=0.9\linewidth]{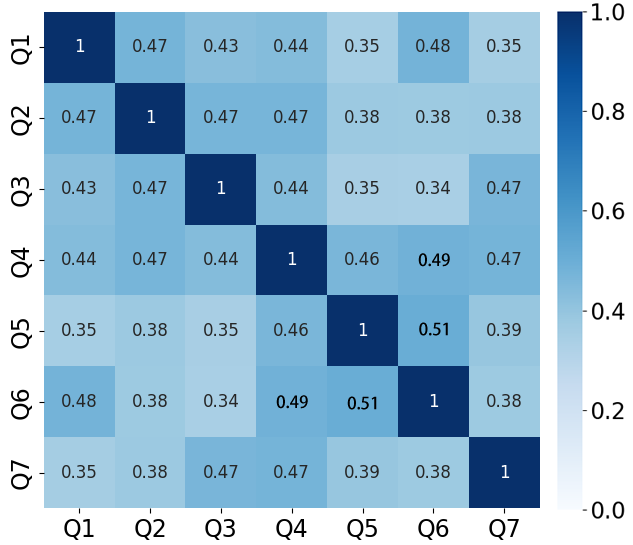}
    \caption{Correlation coefficients for pairs of logical qubits for the quantum Hamming code $[[15, 7, 3]]$. Here $Q_i$ denotes the $i$-th logical qubit. }
    \label{fig:correlation}
\end{figure}
We estimate the marginal probabilities $P(L_i), P(L_j)$ and $P(L_i, L_j)$ using Monte Carlo simulation, and the estimated values for $\rho_{n_i, n_j}$ are shown in Fig.~\ref{fig:correlation}. It is evident that the correlation coefficients for all pairs of logical qubits are positive and are around $0.4$, indicating a significant amount of correlation. The positive correlation implies that if a logical qubit does not suffer an error, the other logical qubits remain error free with a high probability.

The correlation exists not only between a pair of qubits, but also within a larger set of qubits (see Appendix~\ref{app:LSC} for more details). Furthermore, we find numerical evidence suggesting that the logical errors are locally decaying (see Appendix~\ref{app:LDE} for more details), which is a crucial assumption needed to prove the existence of an error threshold~\cite{Yamasaki_2024}.

\subsection{Error threshold}

\begin{figure}[htbp]
    \centering
    \includegraphics[width=0.95\linewidth]{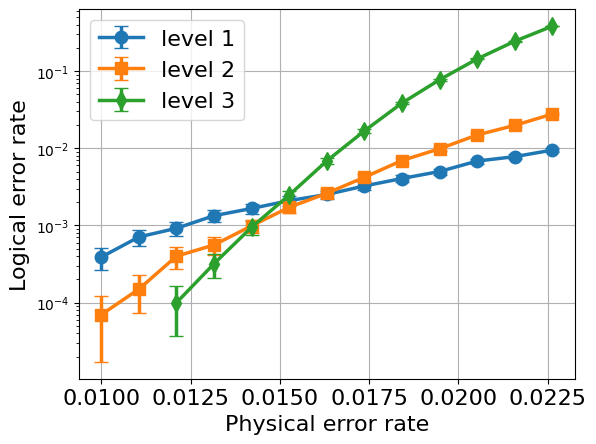}
    \caption{ Logical error rate of concatenated quantum Hamming codes  that are concatenated up to level 1, level 2 and level 3. The threshold is about $p_{th} \approx 1.5 \%$. The error bars are calculated at the 95\% confidence interval.}
    \label{fig:cssl}
\end{figure}

In order to estimate the threshold, we simulate the decoding process on three concatenated quantum Hamming codes, which are concatenated up to level 1, level 2, level 3, and 
correspond to the $[[15, 7, 3]]$, $[[31, 21, 3]]$, $[[63 ,51, 3]]$ quantum Hamming codes, respectively. 
In simulation of the decoding process, we randomly generate independent errors in level-0 registers (physical qubits) with a physical error rate $p$, apply the decoding algorithm and correction operators, and check if each logical qubit has suffered a logical error. If any logical qubit is not error-free, we count the result as a logical error. The logical error rate is estimated by repeating the above simulation multiple times. Figure~\ref{fig:cssl} shows the estimated logical error rate as a function of the physical error rate $p$, with each simulation repeated $10^5$ times.

In Ref.~\cite{yoshida2024concatenate}, instead of simulating the entire decoding process for the concatenated quantum Hamming code, the quantum Hamming codes at each level are decoded separately, and the relation between the logical error rate and the physical error rate is derived at each level. In order to obtain the logical error rate of the full concatenated quantum Hamming code, one substitutes the logical error rate at level $l$ as the physical error rate of level $l+1$ and continues this calculation iteratively using the relation between the logical error rate and physical error rate at each level. while in our work, we simulate the entire decoding process.

During the decoding process, the error correction at each level may fail, resulting in logical errors at that level. As observed in Sec.~\ref{sec:corrlated error}, the output for each level are noisy logical qubits with correlated errors. 
Our comprehensive numerical simulation demonstrates that an error threshold exists, albeit with correlated logical errors at each level of concatenation. From Fig.~\ref{fig:cssl} it can be seen that the threshold is about $p_{th} \approx 1.5\%$. The current NISQ devices can achieve Pauli error rates for a single qubit of approximately 0.5\%~\cite{dasilva2024demonstrationlogicalqubitsrepeated} and 0.8\%~\cite{acharya2024quantumerrorcorrectionsurface}.
Therefore, the concatenated quantum Hamming codes provide a promising candidate for a quantum memory.

\section{Concatenation of Surface codes with quantum Hamming codes}\label{sec:surface-Hamming}

In this section, we consider a variant of the concatenated quantum Hamming code. The goal is to increase the error threshold to alleviate the difficulty of experimental implementation. 
Since the surface code has a relatively higher error threshold, we concatenate it with quantum Hamming codes in order to increase the error threshold. 
The strategy is to replace the level-$0$ code, namely the $[[7, 1, 3]]$ code, by a family of surface codes and keep the higher levels of concatenation the same as that of the concatenated quantum Hamming codes. By increasing the size of the surface code, its logical errors can be exponentially suppressed. It is therefore expected that a higher threshold can be achieved by concatenating a surface code of a larger size with quantum Hamming codes. 


The decoding algorithms for the surface codes and quantum Hamming codes are different, therefore a complete decoding process is divided into two steps. Firstly, the surface codes in the level-$0$ encoding are decoded using standard decoding algorithms, e.g., minimum-weight-perfect-matching algorithm. The outputs are noisy registers with independent errors and are then fed into the next level of concatenation with quantum Hamming codes. Secondly, the quantum Hamming codes are decoded using the decoder proposed in Sec.~\ref{sec:decoder}. 

To simplify the decoding process and conserve computational resources in the simulation, we decode only one block of surface code for a given size and estimate the relation between the logical error rate and physical error rate. Independent logical Pauli errors are then randomly generated in all blocks of surface code following the probability distribution based on the estimated logical error rate. These noisy blocks of surface code are then concatenated with quantum Hamming codes and are decoded accordingly.

\begin{widetext}
    
\begin{figure*}
\subfigure[] {\includegraphics[width=0.32\columnwidth]{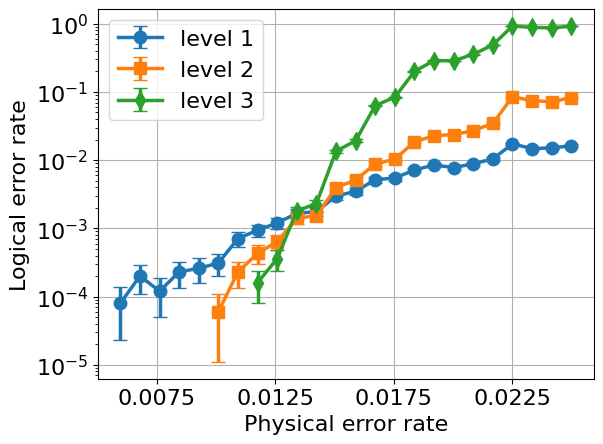}}
\subfigure[]
{\includegraphics[width=0.32\columnwidth]{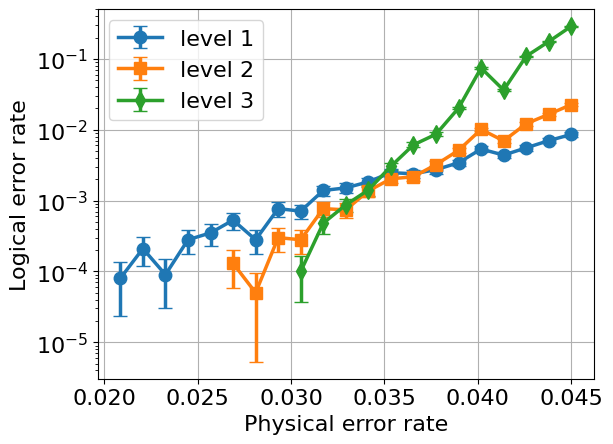}}
\subfigure[]
{\includegraphics[width=0.32\columnwidth]{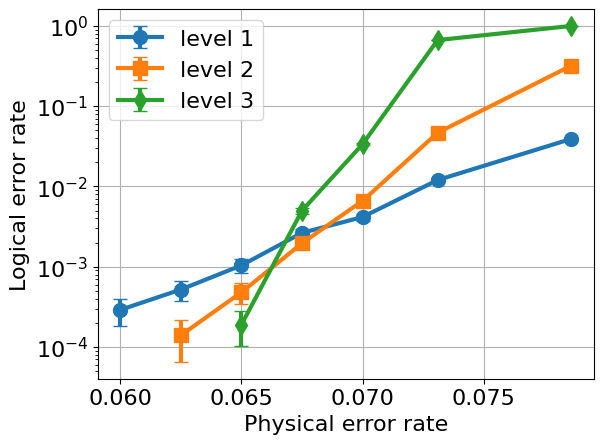}}
\caption{Error thresholds for surface-Hamming codes. The code distance $d$ of the underlying surface code is (a) $d=3$, (b) $d = 5$ and (c) $d = 21$. The error bars are calculated at the 95\% confidence interval.}
\label{fig:surface-hamming}
\end{figure*}

\end{widetext}

The logical error rates of surface-Hamming codes are estimated using Monte-Carlo simulation and the results are shown in Fig.~\ref{fig:surface-hamming}. It is evident that an error threshold exists for the surface-Hamming code built upon the surface code of any given size. In particular, when the physical error rate is above the error threshold, the logical error rate increases with more levels of concatenation. Conversely, when the physical error rate is below the error threshold, the logical error rate decreases with more levels of concatenation. One can also see from Fig.~\ref{fig:surface-hamming} that the error threshold increases by concatenating a surface code with a larger size. The error threshold is about $1.3\%$ for the surface code with $d=3$ and increases up to about $3.4\%$ for $d=5$, and above $6.5\%$ for $d=21$. The logical error rate at threshold remains almost the same, which is about $10^{-3}$. By further increasing the size of the surface code, it is expected that the error threshold can approach to the threshold of the surface code. 

An important quantity that characterizes the performance of a quantum code is its resource overhead, which is defined as the number of physical qubits required to encode a single logical qubit. It can also be considered as the inverse of the code rate. The resource overhead for the concatenation of the surface code with the quantum Hamming code up to different levels are listed in Table~\ref{tab:Rate}. The resource overhead for the level-$0$ concatenation is simply the overhead for surface codes, which is given by $d^2 + (d-1)^2$. Note that concatenating a smaller surface code with quantum Hamming codes up to high levels may have a comparable resource overhead with a larger surface code. For example, concatenating a surface code with $d=3$ with quantum Hamming codes up to level-$2$ has almost the same overhead as a surface code of size $d=5$. 

\begin{table}
    \centering
    \caption{ Resource overhead of the surface-Hamming codes that are concatenated up to different levels. Here $d$ is the code size of the underlying surface code. The resource overhead in the first column for level-0 concatenation is the overhead for the surface code. 
    }
    \begin{tabular}{c|c|c|c|c}
        \hline \hline
         {\bf Code distance} & ~{\bf Level 0}~ & ~{\bf Level 1}~  & ~{\bf Level 2}~ & ~{\bf Level 3}~ \\ \hline
       $d=3$ & 13  & 28 & 41 & 51 \\ \hline
      $d=4$ & 25  & 54  & 79  & 98 \\ \hline
      $d=5$ & 41  & 88  & 130  & 160 \\ 
      \hline
    \end{tabular}
    \label{tab:Rate}
\end{table}

\begin{widetext}



    
\begin{figure}[htbp]
    \centering
    \includegraphics[width=0.9\linewidth]{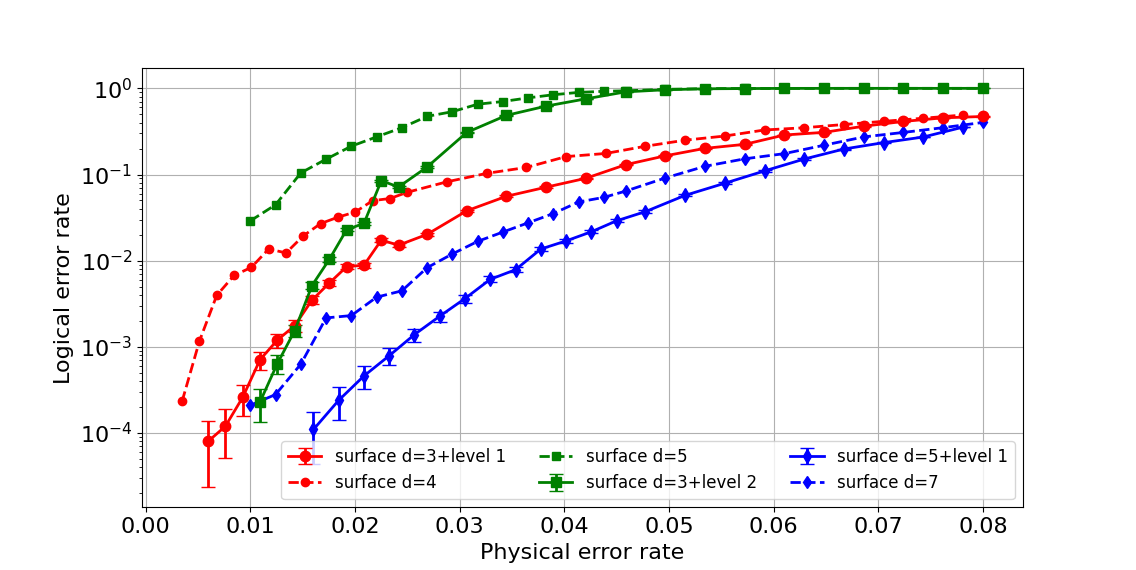}
    \caption{Comparison of logical error rates between surface-Hamming codes and surface codes without concatenation, given a comparable amount of resource overhead. The logical error rate of the surface code is calculated by $1-(1-P_{\rm sur})^k$, where $P_{\rm sur}$ is the logical error rate of a single block of surface code, and $k$ is the total number of logical qubits encoded in a quantum memory. Note that curves with the same color represent the logical error rates calculated from the same number of logical qubits. The error bars are calculated at the 95\% confidence interval.}
    \label{fig:rate-comparison}
\end{figure}

\end{widetext}

We compare the performance of two encoding schemes for implementing a multi-qubit quantum memory: one using the surface-Hamming codes and the other using multiple blocks of surface codes. In particular, we compare their logical error rates under the condition that they consume a comparable amount of resource overhead. This comparison is crucial for understanding the efficiency and practicality of different quantum error correction codes when implemented under similar constraints. The one that achieves a lower logical error rate would be more favorable for experimental implementation. From Table~\ref{tab:Rate} we can see that the overhead for concatenating a surface code of size $d=3$ with quantum Hamming codes up to level $1$ is about $28$, which is comparable to the overhead of a surface code with $d=4$. The corresponding relation between the logical error rate and physical error rate is shown in Fig.~\ref{fig:rate-comparison}. It is evident that the logical error rate for the surface-Hamming code is lower than that of the surface code without concatenation when the physical error rate is below $7\,\%$. 
The gap between the logical error rates of these two encoding schemes is more pronounced for a lower physical error rate. For instance, the logical error rate of the surface-Hamming code is about an order of magnitude lower when the physical error rate is around $1\,\%$. Similar behaviors can also be observed by comparing the logical error rates of the concatenation of the surface code of size $d=3$ with quantum Hamming codes up to level $2$ and the surface code of size $d=5$ without concatenation, and the concatenation of surface code of size $d=5$ with quantum hamming codes up to level $1$ and the surface code of size $d=7$ without concatenation.
These results demonstrate that concatenating surface codes with quantum Hamming codes is more powerful in suppressing logical errors than the surface codes alone, given a comparable amount of resource overhead.

We also find
that concatenating the surface code of size $d=3$ with quantum Hamming codes up to level $2$ can outperform the surface code of size $d=7$ without concatenation. Specifically, the logical error rate of the former is lower than that of the latter when the physical error rate is below $2 \,\%$. The resource overhead of the latter is $85$, while the overhead of the former is about $41$, saving roughly half the overhead. This demonstrates that concatenating surface codes with quantum Hamming codes up to higher levels can achieve better performance in suppressing logical errors while consuming less resource overhead. 

Interestingly, the surface-Hamming code starts to show advantages for a quantum memory at an intermediate scale, but not in the asymptotic limit. Using blocks of surface code of size $d=3$, seven logical qubits are encoded up to level $1$, requiring a total of about $195$ physical qubits; $147$ logical qubits are encoded up to level $2$, requiring a total of about $6027$ physical qubits. 
Quantum memories of this intermediate scale can potentially be achieved in the near future. 



\section{Decoding time}\label{sec:decoding-time}

\begin{figure}[htbp]
\includegraphics[width=0.9\columnwidth]{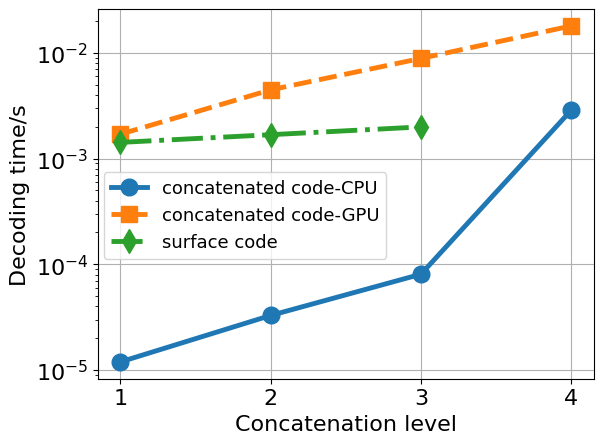}
\caption{Comparison of decoding time for concatenated quantum Hamming codes and surface codes. }
\label{fig:time}
\end{figure}

In this section, we estimate the decoding time of the concatenated quantum Hamming codes. 
Note that the time spent on correction operation is not included 
because it depends on the specific hardware implementation. 
To reduce the effect of runtime fluctuation, we repeat the decoding process for $10^5$ times and calculate the average runtime. We estimate the decoding time by running the algorithm on both the GPU and CPU, 
with parallel computing on the GPU and no parallel computing on the CPU. The computational devices we used are Intel Xeon Silver 4214R CPU and RTX A6000 GPU.
Due to the lack of parallel computing capabilities on the CPU, we must decode each code block sequentially at every level. Given that the number of code blocks grows exponentially with the number of concatenation, the algorithm executed on the CPU exhibits $O(\exp({\rm poly}(\ell)))$ time complexity. As shown in Fig.~\ref{fig:time}, the blue thick line demonstrates this exponential growth. In contrast, the GPU can decode all code blocks simultaneously using parallel computing, achieving a lower growth rate compared to the CPU execution, as shown in Fig.~\ref{fig:time}.
 Due to the latency caused by loading data into and reading data from the GPU, the GPU execution has an increased computation time than CPU execution.
 To address the latency issue, one potential solution is to design specialized chips that mitigate delays caused by data loading from memory into the GPU.

The algorithm we used for decoding the surface code is the minimum weight perfect matching, which is implemented in Python by Pymatching~\cite{higgott2021pymatchingpythonpackagedecoding}, and executed on CPU. 
We compare the decoding time of the surface code and that of the concatenated quantum Hamming code, ensuring both achieve a comparable logical error rate at a physical error rate of $p=0.01526$. We find that a surface code with at least $d=5$ is needed to achieve a comparable logical error rate to a level-1 concatenated quantum Hamming code. For a level-2 concatenated code, a surface code with at least $d=7$ is required; and for a level-3 concatenated code, a surface code with at lease $d=9$ is required. From Fig.~\ref{fig:time} it can be observed that when the number of concatenation is less than 4, the concatenated quantum Hamming code exhibits a computational advantage on the CPU. The empirically determined runtime of Pymatching is $O(D^{2.1})$~\cite{higgott2021pymatchingpythonpackagedecoding}, while our algorithm achieves a runtime of $O(\exp({\rm poly}(\ell)))$ on the CPU without parallel computation. Despite the polynomial scaling of runtime of Pymatching and exponential scaling of runtime of our algorithm, our algorithm exhibits computational advantages on the same hardware. This advantage is likely due to the smaller coefficients in the exponential scaling of the concatenated quantum Hamming code, which becomes significant for smaller values of $\ell$.

\section{Conclusions}\label{sec:conclusion}

We conducted a complete numerical simulation of the decoding process and estimated the error threshold of a multi-qubit quantum memory based on the concatenated quantum Hamming codes and the surface-Hamming codes, under the assumption of perfect syndrome extraction and measurement. The complete numerical simulation allows us to take into account correlated errors that occur during the error correction at each level. Our numerical results show that an error threshold exists for the concatenated quantum Hamming code. By concatenating surface codes with quantum Hamming codes, the error threshold can be significantly improved, and it approaches to the threshold of surface code when increasing the size of the surface code in the lowest level of concatenation. Interestingly, the surface-Hamming code outperforms the surface code without concatenation in suppressing logical errors, assuming a comparable amount of resource overhead. This advantage of the surface-Hamming code appears for a quantum memory of intermediate size, making it more favorable for experimental realization in the near future. 

When neglecting errors from syndrome extraction and measurement, our study only reveals intrinsic properties of the surface-Hamming code. The high error threshold, low resource overhead, fast decoder and good performance in suppressing logical errors make it a very competitive candidate for fault-tolerant quantum computation. However, it remains an open problem whether these advantages still survive if the errors from syndrome extraction and measurement are taken into account. The quantum Hamming codes are not quantum LDPC codes, implying that they may require a very deep syndrome extraction circuit. Errors accumulated in the syndrome extraction circuit may significantly reduce the error threshold, negating any advantages. The study in Ref.~\cite{yoshida2024concatenate} indicates that this may not be the case. We leave the study that takes into account more realistic scenarios for future work. 

{\bf Acknowledgements:} D. S. is supported by the Fundamental Research Funds for the Central Universities, HUST (Grant No. 5003012068) and Wuhan Young Talent Research Funds (Grant No. 0106012013). The numerical simulation in this paper is performed in the HPC platform of Huazhong University of Science and Technology. 

\appendix
\vspace{0.5cm}


\section{Error probability for individual logical qubits} 
\label{sec::error rate}

\begin{figure}[htbp]
    \centering
    \includegraphics[width=0.95\linewidth]{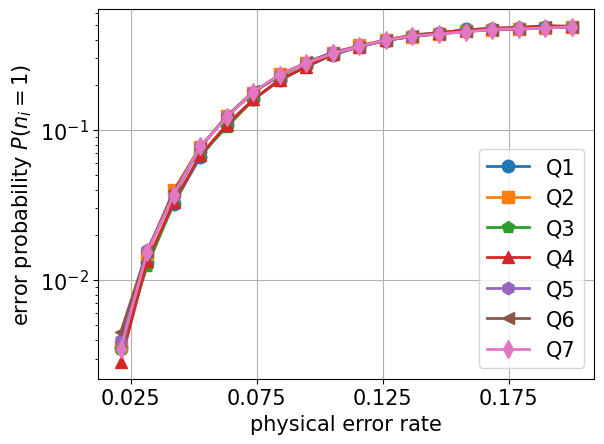}
    \caption{Estimated error probability of different logical qubits for the $[[15, 7, 3]]$ quantum Hamming code with a physical error rate $p=0.08$. We performed $20000$ rounds of simulation and count the logical errors to estimate the error probability. 
    }
    \label{fig:indepen}
\end{figure}

In the simulation, the error probability of each logical qubit for the [[15,7,3]] quantum Hamming code is computed, as  illustrated in Fig~\ref{fig:indepen}. The results indicate that the error probabilities of the logical qubits are nearly identical.

\section{Locally decaying errors}
\label{app:LDE}

When errors are correlated, threshold still exists if the errors are locally decaying~\cite{pattison2023hierarchical}.
To define \textit{locally decaying error}, let $E_S$ be any set containing the errors occurring on physical qubits, where the elements of $E_S$ are single-qubit errors. For example, $X_1X_2$ would be represented by $E_S=\{X_1,X_2\}$. $|E_S|$ is the cardinal number of $E_S$. $P(E_S)$ is the marginal probability distribution of the event that $E_S$ has occured. The definition of \textit{locally decaying error} is as follows.
\begin{definition}
 If all selections of $E_S$ and $P(E_S)$ satisfy
    \begin{equation}
        P(E_S)\leq p^{|E_S|},
    \end{equation}
    where $p$ is a constant, then the distribution $P(E_S)$ is locally decaying with rate $p$, and the error model generating $E_S$ is a \textit{locally decaying error} model.
\end{definition}

\begin{figure}[htbp]
    \centering
    \includegraphics[width=0.95\linewidth]{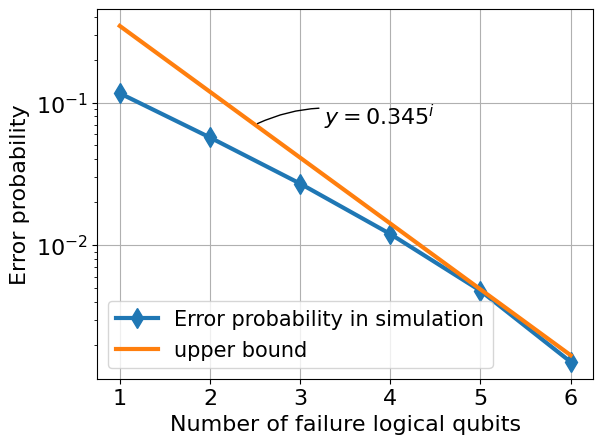}
    \caption{
    Error probability for different number of failure logical qubits and its upper bound. The blue line represents $p_i$ and the yellow line represents the upper bound of $p_i$, given by $y=0.345^i$. 
  }
    \label{fig:decay}
\end{figure}

From Appendix~\ref{sec::error rate}, we know that the error probability of each logical qubit is almost identical when $|E_S|=1$. We assume that this also holds when $|E_S|>1$. In particular, we assume that the probability of errors with equal cardinal number is approximately identical, namely, 
\begin{equation}
    P(E_S)=p_i
\end{equation}
for all $E_S$ satisfying $|E_S|=i$, where $i\in[1,K]$ with $K$ the number of logical qubits. Therefore, under this assumption, we only need to verify whether $p_i\leq p^i$ for a given $p$. By counting the errors in $20000$ rounds of simulation in Appendix~\ref{sec::error rate}, we obtained the error probability, as shown in Fig.~\ref{fig:decay}.
From Fig.~\ref{fig:decay} we can find that $p_i<0.345^i$. Therefore, the logical error rate of the $[[15, 7, 3]]$ quantum Hamming code is locally decaying with rate $0.345$. The conclusion that the logical error rate of the $[[15, 7, 3]]$ quantum Hamming code is locally decaying also holds for other physical error rates. 

While we only demonstrate a locally decaying error model for the
$[[15, 7, 3]]$ quantum Hamming code, 
we expect that this also holds for 
other quantum Hamming codes. Based on this indication, threshold can exist for concatenated quantum Hamming codes even when there exists correlations within logical qubits.

\section{Large set correlations}
\label{app:LSC}

\begin{widetext}
    
\begin{figure*}[htbp]
\subfigure[] {\includegraphics[width=0.445\columnwidth]{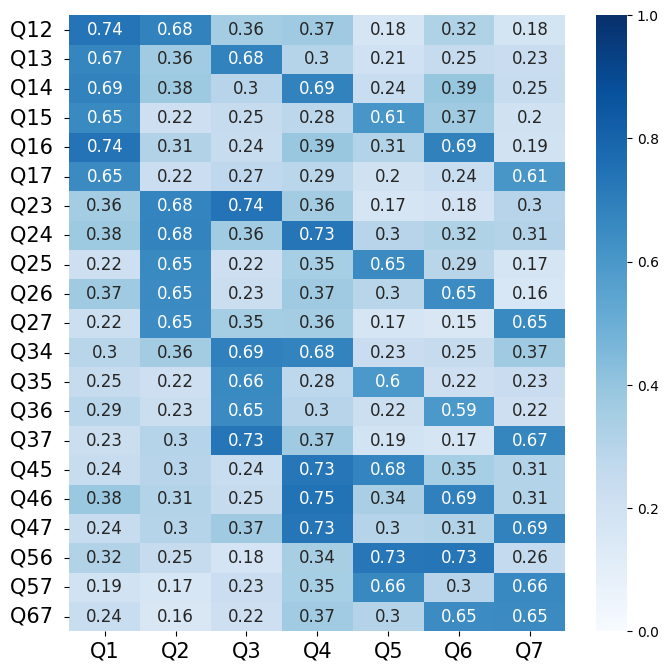}}
\subfigure[]
{\includegraphics[width=0.447\columnwidth]{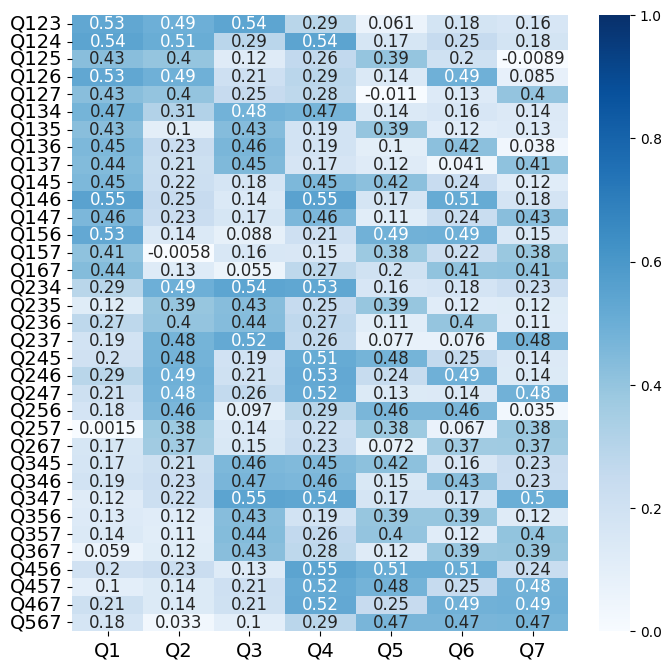}}\\
\subfigure[]
{\includegraphics[width=0.44\columnwidth]{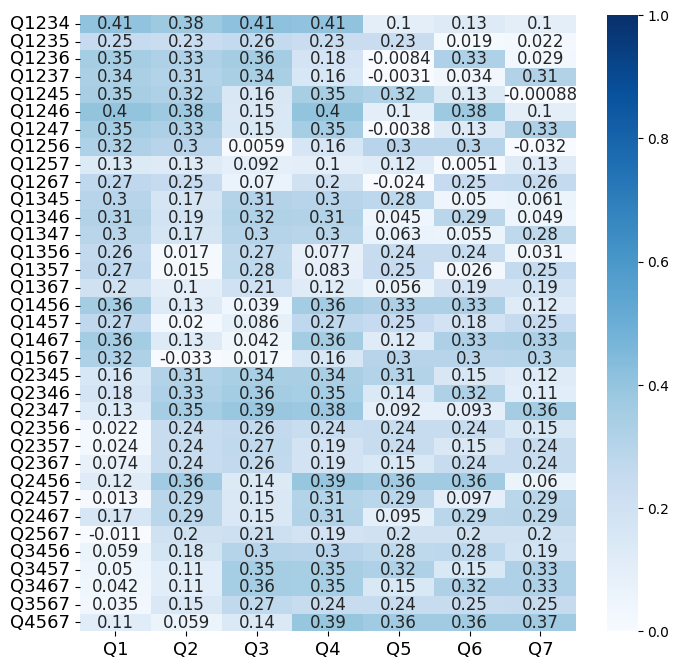}}
\subfigure[]
{\includegraphics[width=0.458\columnwidth]{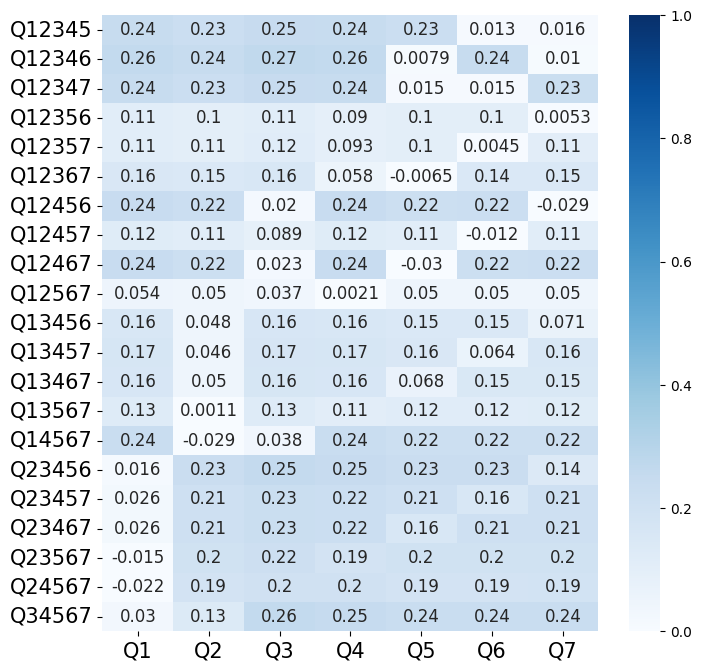}}
\caption{Correlation coefficients for a set of logical qubits and individual logical qubits. These sets are (a) 2-qubit sets, (b) 3-qubit sets, (c) 4-qubit sets and (d) 5-qubit sets.}
\label{fig:corrsets}
\end{figure*}

\end{widetext}

We also calculate the correlation coefficients between a large set of logical qubits and individual logical qubits, and the results are shown in Fig.~\ref{fig:corrsets}. Note that the $y$-axis is labeled as a set of logical qubits, while the $x$-axis is labeled for an individual logical qubit. 
A logical error is counted and labeled as 1 when every logical qubit in the set has suffered an error; while for all other cases, the event is labeled as 0. We performed 20000 rounds of simulation and estimated the correlation coefficients between a set of logical qubits and individual logical qubits.
From Fig.~\ref{fig:corrsets} we can find that the correlations  between a set of logical qubits and individual logical qubits are positive for most cases. 

\clearpage

\section{Logical operators for $n=5$}
\label{app:logical-operator}

The matrix characterizing the logical operators of the quantum Hamming code $[[31, 21, 3]]$ is given by

\begin{widetext}
\begin{equation}
L_5=\left(             
  \begin{array}{ccccccccccccccccccccccccccccccc}   
1 &0 &0 &0 &0 &0 &0 &0 &0 &0 &0 &0 &0 &0 &0 &0 &0 &0 &0 &0 &0 &0 &0 &0 &0 &0 &0 &0 &0 &1 &1 \\
0 &1 &0 &1 &0 &0 &0 &1 &0 &0 &0 &0 &0 &0 &0 &1 &0 &0 &0 &0 &0 &0 &0 &0 &0 &0 &0 &0 &0 &1 &0 \\
1 &1 &1 &0 &0 &0 &0 &0 &0 &0 &0 &0 &0 &0 &0 &0 &0 &0 &0 &0 &0 &0 &0 &0 &0 &0 &0 &0 &0 &0 &0 \\
1 &0 &0 &1 &0 &0 &0 &1 &0 &0 &0 &0 &0 &0 &0 &1 &0 &0 &0 &0 &0 &0 &0 &0 &0 &0 &0 &0 &1 &0 &0 \\
0 &1 &0 &1 &1 &0 &0 &0 &0 &0 &0 &0 &0 &0 &0 &0 &0 &0 &0 &0 &0 &0 &0 &0 &0 &0 &0 &0 &1 &1 &0 \\
0 &0 &1 &0 &1 &1 &0 &0 &0 &0 &0 &0 &0 &0 &0 &0 &0 &0 &0 &0 &0 &0 &0 &0 &0 &0 &0 &0 &0 &0 &0 \\
0 &1 &0 &1 &0 &0 &1 &0 &0 &0 &0 &0 &0 &0 &0 &0 &0 &0 &0 &0 &0 &0 &0 &0 &0 &0 &0 &0 &0 &1 &1 \\
0 &1 &0 &1 &1 &1 &1 &0 &0 &0 &0 &0 &0 &0 &0 &0 &0 &0 &0 &0 &0 &0 &0 &0 &0 &0 &0 &1 &0 &1 &0 \\
0 &1 &0 &0 &0 &0 &1 &0 &1 &0 &0 &0 &0 &0 &0 &1 &0 &0 &0 &0 &0 &0 &0 &0 &0 &0 &0 &0 &1 &1 &1 \\
0 &1 &0 &1 &1 &1 &0 &1 &1 &1 &0 &0 &0 &0 &0 &1 &0 &0 &0 &0 &0 &0 &0 &0 &0 &0 &0 &0 &0 &1 &0 \\
0 &0 &1 &1 &1 &0 &0 &1 &1 &1 &1 &0 &0 &0 &0 &0 &0 &0 &0 &0 &0 &0 &0 &0 &0 &0 &0 &0 &1 &0 &1 \\
0 &1 &1 &1 &1 &0 &1 &0 &1 &0 &0 &1 &0 &0 &0 &0 &0 &0 &0 &0 &0 &0 &0 &0 &0 &0 &0 &1 &0 &1 &0\\
0 &1 &0 &1 &0 &0 &0 &1 &0 &1 &1 &1 &1 &0 &0 &1 &0 &0 &0 &0 &0 &0 &0 &0 &0 &0 &0 &0 &0 &1 &0 \\
0 &0 &0 &1 &1 &1 &1 &0 &0 &0 &0 &0 &0 &1 &0 &1 &0 &0 &0 &0 &0 &0 &0 &0 &0 &0 &0 &1 &1 &0 &1 \\
0 &0 &0 &1 &1 &1 &1 &0 &0 &1 &1 &0 &0 &0 &1 &1 &0 &0 &0 &0 &0 &0 &0 &0 &0 &0 &0 &1 &1 &0 &1 \\
0 &1 &0 &1 &1 &1 &0 &1 &0 &0 &1 &1 &1 &1 &0 &1 &0 &0 &0 &0 &0 &0 &0 &0 &0 &0 &1 &1 &0 &1 &0 \\
0 &0 &1 &0 &1 &0 &1 &1 &1 &0 &0 &0 &0 &1 &1 &1 &1 &0 &0 &0 &0 &0 &0 &0 &0 &0 &0 &0 &0 &0 &0 \\
0 &0 &0 &1 &1 &1 &1 &0 &0 &0 &0 &0 &1 &0 &0 &1 &1 &1 &0 &0 &0 &0 &0 &0 &0 &0 &0 &1 &1 &0 &1 \\
0 &1 &1 &1 &0 &1 &1 &0 &1 &0 &1 &1 &0 &0 &1 &0 &0 &0 &1 &1 &0 &0 &0 &0 &0 &0 &0 &1 &0 &1 &0 \\
0 &1 &1 &1 &1 &0 &1 &0 &1 &1 &1 &0 &1 &1 &1 &0 &0 &0 &1 &0 &1 &0 &0 &0 &0 &0 &1 &0 &0 &1 &0 \\
0 &1 &1 &0 &1 &0 &0 &0 &1 &1 &0 &1 &0 &0 &1 &1 &1 &1 &1 &0 &1 &1 &0 &0 &0 &0 &1 &0 &1 &1 &1 
  \end{array}\right).
\end{equation}
\end{widetext}

\section{Comparison between correlated and independent error models}

To investigate the effect of correlated errors, we compare the logical error rates in the absence and presence of correlated errors.
In the former case, we analyze the logical error rates of the $[[2^{l+3}-1,\ 2^{l+3}-2(l+3)-1,\ 3]]$ quantum Hamming codes for various values of $l$. Since all quantum Hamming codes have code distance three, they can therefore correct only single-qubit errors. The logical error rate can be well approximated by the probability of having errors on at least two qubits, namely, 
\begin{equation}\label{eq:LEHamming}
    P_l(p)=1-(1-p)^{N_l}-N_l(1-p)^{N_l-1}p,
\end{equation}
where $N_l=2^{l+3}-1$ is the total number of qubits of the corresponding quantum Hamming code, $p$ represents the physical error rate. To justify the validity of Eq.~\eqref{eq:LEHamming}, we performed a numerical simulation of the decoding process (for $l=1, 2, 3$) and estimated the relation between the logical error rate and the physical error rate, and found that the numerical result agrees very well with Eq.~\eqref{eq:LEHamming}. 

We then calculate the logical error rates for the concatenated quantum Hamming codes using  the following iterative procedure,
\begin{equation}
    P_{L}\circ P_{L-1} \circ P_{L-2}\circ \cdots P_0(p),
\end{equation}
where $P_j \circ P_{j-1}(p)$ represents a composition $P_j(P_{j-1}(p))$. Note that the above calculation assumes the absence of correlated errors within every concatenation level. This implies that when the logical error rate of the current level is used as the physical error rate for the next level, it is assumed that each physical qubit in the next level is independent. However, if the errors are correlated, the assumption of independence must be violated.

\begin{figure}[htbp]
    \centering
    \includegraphics[width=0.95\linewidth]{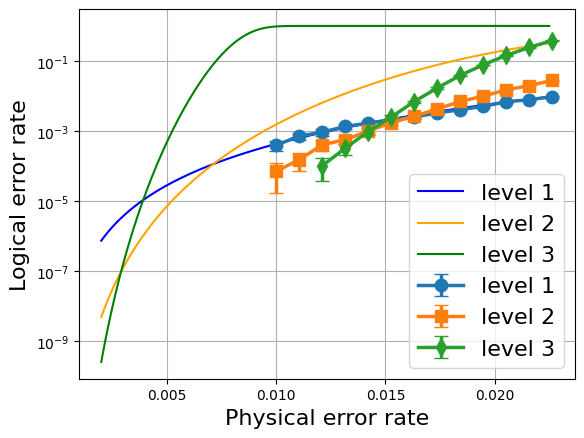}
    \caption{Logical error rate for concatenated quantum Hamming codes. The solid lines represent the logical error rate derived from assuming independent errors, while the dotted lines represent the logical error rate derived from a full decoding processing that takes into account correlated errors.}
    \label{fig:appdix}
\end{figure}

We compare the logical error rates derived from assuming the absence and presence of correlated errors, as shown in Fig.~\ref{fig:appdix}. Note that the logical error rate at level $1$ is the same for both cases. This is because the $[[7, 1, 3]]$ quantum code at level $0$ encodes only a single logical qubit, giving rise to independent errors at the level-$1$ concatenation. However, for higher levels of concatenation, taking into account the correlated errors significantly reduces the logical error rate and increases the error threshold. This shows that correlated errors have to be taken into account in practical decoding.

\bibliography{surface-hamming}

\end{document}